\documentclass[singlecolumn,showpacs,preprintnumbers,amsmath,amssymb]{revtex4}
\usepackage{graphicx}% Include figure files
\usepackage{dcolumn}% Align table columns on decimal point
\usepackage{bm}% bold math

\begin{document}
%\preprint{APS/123-QED}

\title{Role of surface energy coefficients and nuclear surface diffuseness in the fusion of heavy-ions}
% Force line breaks with \\

\author{Ishwar Dutt }
%\altaffiliation [Also at ]{Physics Department, XYZ University.}%Lines break automatically or can be forced with \\
\author{Rajeev K. Puri}%
\email{rkpuri@pu.ac.in;drrkpuri@gmail.com} \affiliation{Department
of Physics, Panjab University, Chandigarh -160 014, India.}

%\affiliation{

\date{\today}% It is always \today, today,

\begin{abstract}
We discuss the effect of surface energy coefficients as well as
nuclear surface diffuseness in the proximity potential and
ultimately in the fusion of heavy-ions. Here we employ different
versions of surface energy coefficients. Our analysis reveals that
these technical parameters can influence the fusion barriers by a
significant amount. A best set of these parameters is also given
that explains the experimental data nicely.
\end{abstract}


% PACS, the Physics and Astronomy % Classification Scheme.
\maketitle

%break was forced \lowercase{via} \textbackslash\textbackslash}

It is now well accepted that Coulomb interactions alone cannot
define a fusion barrier. Nuclear interactions play an equally
important role in deciding the fate of a
reaction~\cite{blocki77,wr94,ms2000,rkp05,aw95}. This is perhaps
the compelling cause of so many new potentials one sees in the
literature~\cite{blocki77,wr94,ms2000,rkp05,aw95}.
 Among
various nuclear potentials, one has the potentials within the
proximity concept~\cite{blocki77,wr94,ms2000}, as well as within
the energy density formalism~\cite{rkp05}. As many as two dozen
potentials and their different versions are being used in the
literature. It is also evident from the literature that every
author has tried to justify the validity of the potential by
showing that it reproduces the proximity values~\cite{rkp05}. At
the same time, it is also interesting to note that several
 improvements were proposed over the original proximity potential in recent times~\cite{wr94,ms2000}.
 In the original version of the proximity potential (labeled as Prox 77)~\cite{blocki77},
 $V_{N}(r)$
 can be written as;
 \begin{equation}
V_{N}\left(r \right) = 4\pi
\overline{R}\gamma\Phi(r-C_{1}-C_{2}){~\rm MeV}, \label{eq:1}
\end{equation}
where $\Phi(r-C_{1}-C_{2})$ is the universal function that was
derived by several authors~\cite{blocki77,ms2000,rkp05}.
$\overline{R}$ is the reduced radius, and $\gamma$ is the surface
energy coefficient.
 \par
The strength of the nuclear potential depends on the relative
neutron excess of the target/projectile through surface energy
coefficient $\gamma$ and on the mass and surface diffuseness
through the reduced radius $\overline{R}$. Though (in-depth)
attention was paid in the literature to pin down the universal
function $\Phi(r-C_{1}-C_{2})$ accurately~\cite{ms2000}, one takes
a very casual approach toward the surface energy coefficients
$\gamma$ and nuclear surface diffuseness. Almost all
models~\cite{blocki77,wr94,ms2000,rkp05,aw95} have used different
terms and/or values for these coefficients. One wonders how much
these parameters can alter the results of fusion barriers and
cross sections. Furthermore, it was reported that the original
proximity potential overestimates the barriers by an appreciable
amount~\cite{ms2000}. We are here interested in studying the
impact of various surface energy coefficients and nuclear surface
diffuseness on the fusion process and shall present a modified
version of the proximity potential based on a new set of surface
energy coefficients.
\par
In the original proximity potential [Eq. (\ref{eq:1})], $C_{1}$
and $C_{2}$ denote the radii of the spherical target/projectile
and are known as S\"ussmann's central radius. The surface energy
coefficient $\gamma$ was taken from the work of W. D. Myers and
\'Swi\c{a}tecki~\cite{ms66} which reads as;
\begin{equation}
\gamma = \gamma_{0}\left[1-k_{s}A_{s}^{2} \right].
 \label{eq:2}
\end{equation}
Here, $A_{s}=\left(\frac{N-Z}{A}\right)$ where N and Z refer to
the total neutron and proton content. In the above formula, $
\gamma_{0}$ is the surface energy constant and $k_{s}$ is the
surface-asymmetry constant. Both constants were first
parameterized by Myers and \'Swi\c{a}tecki~\cite{ms66} by fitting
the experimental binding energies. The first set of these
constants yielded values $\gamma_{0}$ and $k_{s}=1.01734 ~\rm
 ~MeV/fm^{2}$ and 1.79, respectively. Later on, these
values were  revised to ${~\rm \gamma_{0}}$ = 0.9517 $\rm
 ~MeV/fm^{2}$ and
$ k_{s}=1.7826$~\cite{ms67}.  This value of $\gamma$ is referred
as $\gamma$-MS.
\par
In an another attempt,  M\"oller and Nix~\cite{mn76} fitted the
surface energy coefficient $\gamma$ with the value
$\gamma_{0}=1.460734 {~\rm MeV/fm^{2}}$ and $k_{s}=4.0$ in nuclear
macroscopic energy calculations. Naturally, this will lead to more
attraction compared to $\gamma$-MS. This version of $\gamma$ is
labeled as $\gamma$-MN1976.
\par
 Later on, due to the
availability of a better mass formula  due to M\"oller et
al.~\cite{mn95}, $\gamma_{0}$ and $k_{s}$ were refitted to a
strength of $1.25284 {~\rm MeV/fm^{2}}$ and $2.345$, respectively.
This particular set of values were obtained directly from a
least-squares adjustment to the ground-state masses of 1654 nuclei
ranging from $^{16}$O to $^{263}$106 and fission-barrier
heights~\cite{mn95}. This modified $\gamma$ is labeled as
$\gamma$-MN1995.
\par
In the new proximity potential version~\cite{ms2000}, Myers and
\'Swi\c{a}tecki, chose $\gamma$ that also depends on the neutron
skin of the interacting nuclei. The expression of $\gamma$
obtained from the droplet model reads as;
\begin{equation}
\gamma = 1/(4\pi r^{2}_{0})\left[18.63  {\rm  (MeV)}
-Q\left(t^{2}_{1} + t^{2}_{2}\right)/2r^{2}_{0} \right],
\label{eq:3}
\end{equation}
where $t_{i}$ is the neutron skin~\cite{ms2000}. This version of
the surface energy coefficient  is labeled as $\gamma$-MSNew.

\par
As we see, all the previous four versions of $\gamma$ have
different strengths. In fact, in the work of M\"oller and
Nix~\cite{mn81}, as many as five different sets of $\gamma$
parameters were listed. This will, of course, lead to different
values of surface energy coefficients as well as potentials. This
study~\cite{mn81} was based on the calculations of fission-barrier
heights of 28 nuclei and ground-state masses of 1323 nuclei. Royer
and Remaud~\cite{rr84} used  the $\gamma_{0}$ value the same as
$\gamma$-MS with a different $k_{s}$ value (=2.6). Recently,
Pomorski and Dudek~\cite{pd03} obtained different surface energy
coefficients by including different curvature effects in the
liquid drop model. Definitely, the original proximity
potential~\cite{blocki77} used the value of $\gamma$ that was
proposed four decade ago.
\par
 As earlier stated, the nuclear surface diffuseness that enters
 via the reduced radius was also taken in the literature arbitrarily.
 For example, proximity
potentials 1977~\cite{blocki77} and 1988~\cite{wr94} use the
equivalent sharp radius as;
\begin{equation}
R= 1.28A^{1/3}- 0.76+0.8A^{-1/3} {~\rm fm}. \label{eq:5}
\end{equation}
This formula is a semi-empirical expression supported by assuming
the finite compressibility of nuclei. As a result, lighter nuclei
squeeze more by the surface-tension forces, whereas heavy nuclei
dilate more strongly due to the Coulomb repulsion. In both
potentials, the nuclear surface diffuseness enter via reduced
radius $\overline{R}$=$\frac{C_{1}C_{2}}{C_{1}+ C_{2}}$ that is
used in Eq.~(\ref{eq:1}). The central radius $C$ is calculated
from the relation:
\begin{equation}
C-77= R\left[1-(b/R)^{2}+\cdots \cdots \right]. \label{eq:6}
\end{equation}
\par
 For the present study, we
also used the radius due to Aage Winther in Eq.~(\ref{eq:6}) which
reads~\cite{aw95};
\begin{equation}
R= 1.20A^{1/3}-0.09  {~\rm fm}, \label{eq:8}
\end{equation}
and the corresponding central radius~[Eq. (\ref{eq:6})]  is
denoted as C-AW95.
\par
The newer version of the proximity potential uses a different form
of the radius~\cite{ms2000}:
\begin{equation}
R= 1.240A^{1/3} \left\{1+1.646A^{-1}-0.191A_{s}\right\} {~\rm fm}.
\label{eq:9}
\end{equation}
This formula indicates that radius depends not only on the mass
number, it also has a dependence on the relative neutron excess.
Actually this formula is valid for even-even nuclei with $Z \geq
8$~\cite{bn94}. To calculate the matter central radius $C$, the
neutron skin is also added in Ref.~\cite{ms2000} using the
relation
\begin{equation}
C-00= c+ (N/A)t, \label{eq:10}
\end{equation}
where $c$ denotes the half-density radii of the charge
distribution given by
\begin{equation}
c= R[1- (7/2)b^{2}/R^{2}-(49/8)b^{4}/R^{4}+\cdots]. \label{eq:11}
\end{equation}
 Recently, a new form
of Eq.~(\ref{eq:9}) with slightly different constants is also
reported~\cite{gr09}
\begin{eqnarray}
R= 1.2332A^{1/3}+2.8961A^{-2/3}-0.18688A^{1/3} A_{s} .
\label{eq:12}
\end{eqnarray}
%%%%%%%%%%%%%%%%%%%%%%%%%%%%%%%%%%%%%%%%%%%%%%%%%%%%%%%%%%%%%%
 By using this form of the
radius in Eqs.~(\ref{eq:10}) and (\ref{eq:11}), we can again
calculate the central radius $C$ denoted by C-00N.
\par
 Our
calculations are made for 390 reactions involving both the
symmetric N=Z and asymmetric $N\neq Z$ reactions. As noted in
Refs. \cite{blocki77,wr94,ms2000}, the surface energy coefficient
$\gamma$ depends strongly on the asymmetry of the reactions.
\par
In Fig. 1, we display the nuclear potential as a function of
internuclear distance ``r'' for the reactions of $^{12}C+^{12}C$
(in the upper panel) and  $^{6}He+^{238}U$ (in the lower panel)
using the Prox 77 with different versions of surface energy
coefficient $\gamma$. We see that $\gamma$-MS leads to a shallow
potential compared to other sets of $\gamma$, whereas
$\gamma$-MN1976 leads deepest potential. We also tested all the
previously highlighted surface energy coefficients but their value
lies between these extreme limits.
\par
In Fig. 2, we display the fusion barrier heights $V_{B}$ and
fusion barrier positions $R_{B}$ as a function of $Z_{1}Z_{2}$.
For the clarity of the figure, only 155 reactions are displayed.
We show the results of implementing $\gamma$-MS, $\gamma$-MN1976,
$\gamma$-MN1995 and $\gamma$-MSNew as well as different surface
diffuseness in the Prox 77 potential. We see some mild effects in
the outcome. These effects are monotonous in nature. Due to the
wide acceptability of the radius used in Prox 77 (Eq. \ref{eq:5}),
we shall stick to the same formula. The results of different
$\gamma$ values are quantified in Fig. 3, where we display the
percentage deviation over the experimental data. The experimental
data is taken from the
Refs.~\cite{ms2000,Neto90,Gary82,Aguilera86,leigh95,Morsad90,newton01,Vaz81,sikora79,sb81}.
We see that the use of $\gamma$-MS, which is used in the proximity
1977, yield considerable deviations ($\pm 10 \%$). Further, the
use of $\gamma$-MN1976 and $\gamma$-MN1995 yield much improved
results. The average deviations over 390 reactions for the fusion
barrier heights are 3.99\%, 0.77\%, 1.77\%, and 2.37\%, for
$\gamma$-MS, $\gamma$-MN1976, $\gamma$-MN1995, and $\gamma$-MSNew,
respectively. Whereas, for fusion barrier positions, its values
are -1.74\%, 1.95\%, 0.73\%, and 0.0\%, respectively over 272
reactions (barrier positions are not available for all reactions).
\par
 It is clear from the previous study that surface energy coefficients
 $\gamma$-MN1976 and $\gamma$-MN1995 may be better choices. To further strengthen the
 choice we calculate the fusion cross sections using the Wong formula~\cite{wg72}.
\par
In Fig. 4, we display the fusion cross section $\sigma_{fus}$ (in
mb) for the reactions of $^{26}$Mg + $^{30}$Si~\cite{Morsad90},
$^{28}$Si + $^{28}$Si~\cite{Gary82,sb81,Aguilera86}, $^{16}$O +
$^{46}$Ti~\cite{Neto90}, $^{12}$C + $^{92}$Zr~\cite{newton01},
$^{40}$Ca + $^{58}$Ni~\cite{sikora79} and $^{16}$O +
$^{144}$Sm~\cite{leigh95} respectively.
 We see that $\gamma$-MN1976/ $\gamma$-MN1995 give better results
over the original proximity Prox 77.  Note that both fusion
barrier height and curvature affect the sub-barrier fusion
probabilities.
 From the previous analysis, it is
clear that the effect of technical parameters, that is the surface
energy coefficient $\gamma$ as well as surface diffuseness of the
target/ projectile is of the order of 10\%-15\%. The use of
surface energy coefficient $\gamma$-MN1976/ $\gamma$-MN1995
improves the results of the Prox 77 potential considerably. This
modified proximity potential is labeled as ``Proximity 2010''.
\par
In this Brief report, we attempt to understand the role of surface
energy coefficient $\gamma$ as well as nuclear surface diffuseness
in fusion dynamics. Our analysis reveals that these parameters can
affect the nuclear potential as well as fusion barriers by the
same amount as different potentials and one should be careful
while choosing these technical parameters. We also propose a
modified version of Prox 77 with new surface energy coefficient
$\gamma$-MN1976/ $\gamma$-MN1995 which yields closer agreement
with experimental data.

%%%%%%%%%%%%%%%%%%%%%%%%%%%%%%%%%%%%%%%%%%%%%%%%%%%%%%%%%%%%%%%%
 This work was supported by a research grant from the Department of
Atomic Energy, Government of India.

%%%%%%%%%%%%%%%%%%%%%%%%%%%%%%%%%%%%%%%%%%%%%%%%%%%%%
%%%%%%%%%%%%%%%%%%%%%%%%%%%%%%%%%%%%%%%%%%%%%%%%%%%%%%%%%%%%%%%%%%%%%%%%%%%%%%%%%%%%%%%%%%%

%%%%%%%%%%%%%%%%%%%%%%%%%%%%%%%%%%%%%%%%%%%%%%%%%%%%%%%%%%%%%%%%%%%%%
\begin{figure}
\centering
\includegraphics* [scale=0.40] {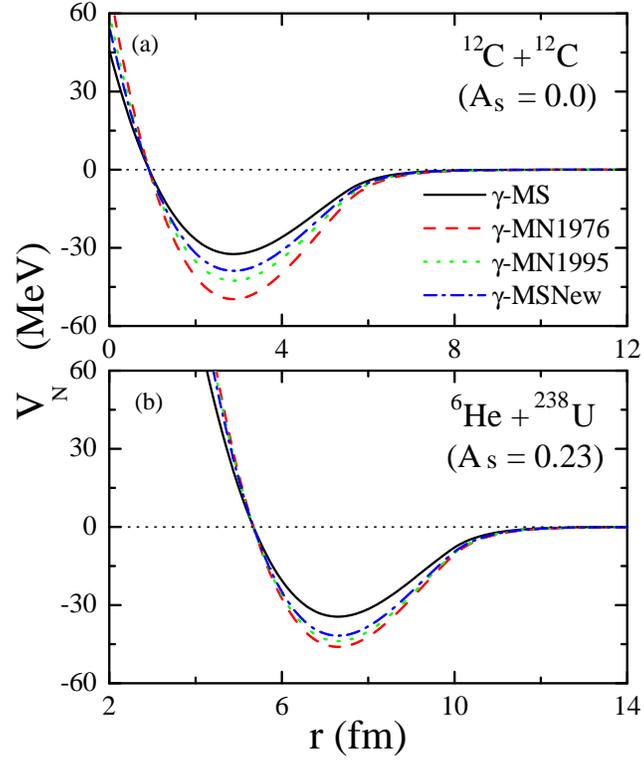}% Here is how to import EPS art
\vskip -0.7 cm \caption {(Color online) The nuclear part
$V_N(MeV)$, of the interaction potential as a function of
internuclear distance ``r'' using Prox 77 with different values of
surface energy coefficients $\gamma$.}
% \vskip -0.7 cm
\end{figure}
%%%%%%%%%%%%%%%%%%%%%%%%%%%%%%%%%%%%%%%%%%%%%%%%%%%%%%%%%%%%%%%%%%%%%
%%%%%%%%%%%%%%%%%%%%%%%%%%%%%%%%%%%%%%%%%%%%%%%%%%%%%%%%%%%%%%%%%%%%%%%%%
\begin{figure}
\centering
\includegraphics* [scale=0.40] {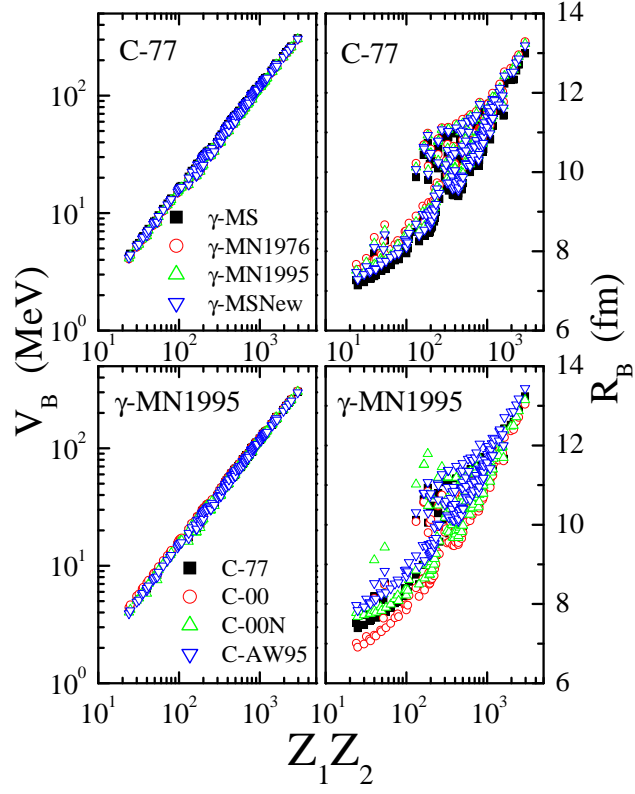}% Here is how to import EPS art
\vskip -0.7cm \caption {(Color online) The fusion barrier heights
$V_{B}$ (MeV) and positions $R_{B}$ (fm) as a function of
$Z_{1}Z_{2}$ using different values of surface energy coefficients
$\gamma$ and nuclear central radii C's implemented in the Prox
77.}
% \vskip-0.6
%cm
\end{figure}
%%%%%%%%%%%%%%%%%%%%%%%%%%%%%%%%%%%%%%%%%%%%%%%%%%%%%%%%%%%%%%%%%%%%%%%%%%
%%%%%%%%%%%%%%%%%%%%%%%%%%%%%%%%%%%%%%%%%%%%%%%%%%%%%%%%%%%%%%%%%%%%%%%%%%%%
\begin{figure}
\centering
\includegraphics* [scale=0.40] {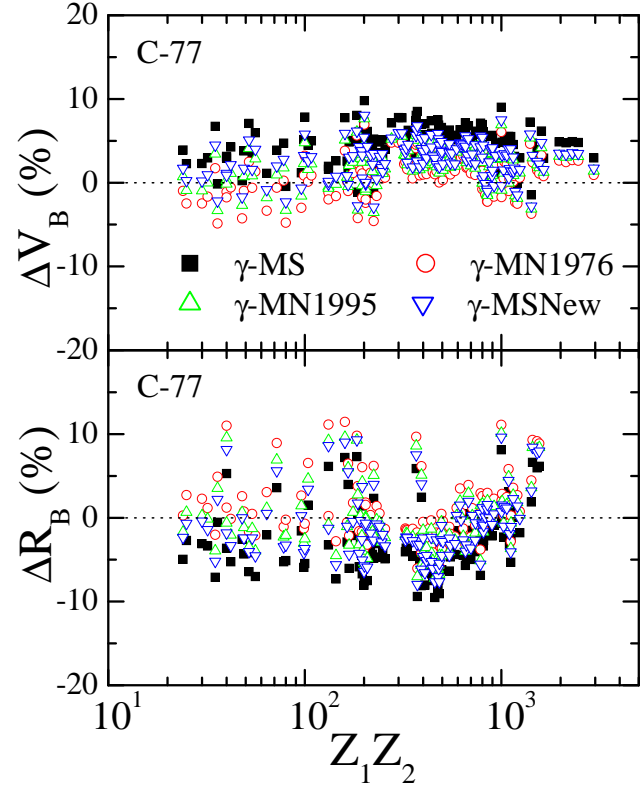}% Here is how to import EPS art
\vskip -0.7 cm \caption {(Color online) The percentage deviation
$\Delta V_{B}~(\%)$ and $\Delta R_{B}~(\%)$ as a function of the
product of charges $Z_{1}Z_{2}$ using different versions of
surface energy coefficients $\gamma$ implemented in the Prox
77.}\label{fig3}
%\vskip -0.6 cm
\end{figure}
%%%%%%%%%%%%%%%%%%%%%%%%%%%%%%%%%%%%%%%%%%%%%%%%%%%%%%%%%%%%%%%
%%%%%%%%%%%%%%%%%%%%%%%%%%%%%%%%%%%%%%%%%%%%%%%%%%%%%%%%%%%%%%%%%%%%%%%%%
\begin{figure}[!t]
\centering
\includegraphics* [scale=0.42]{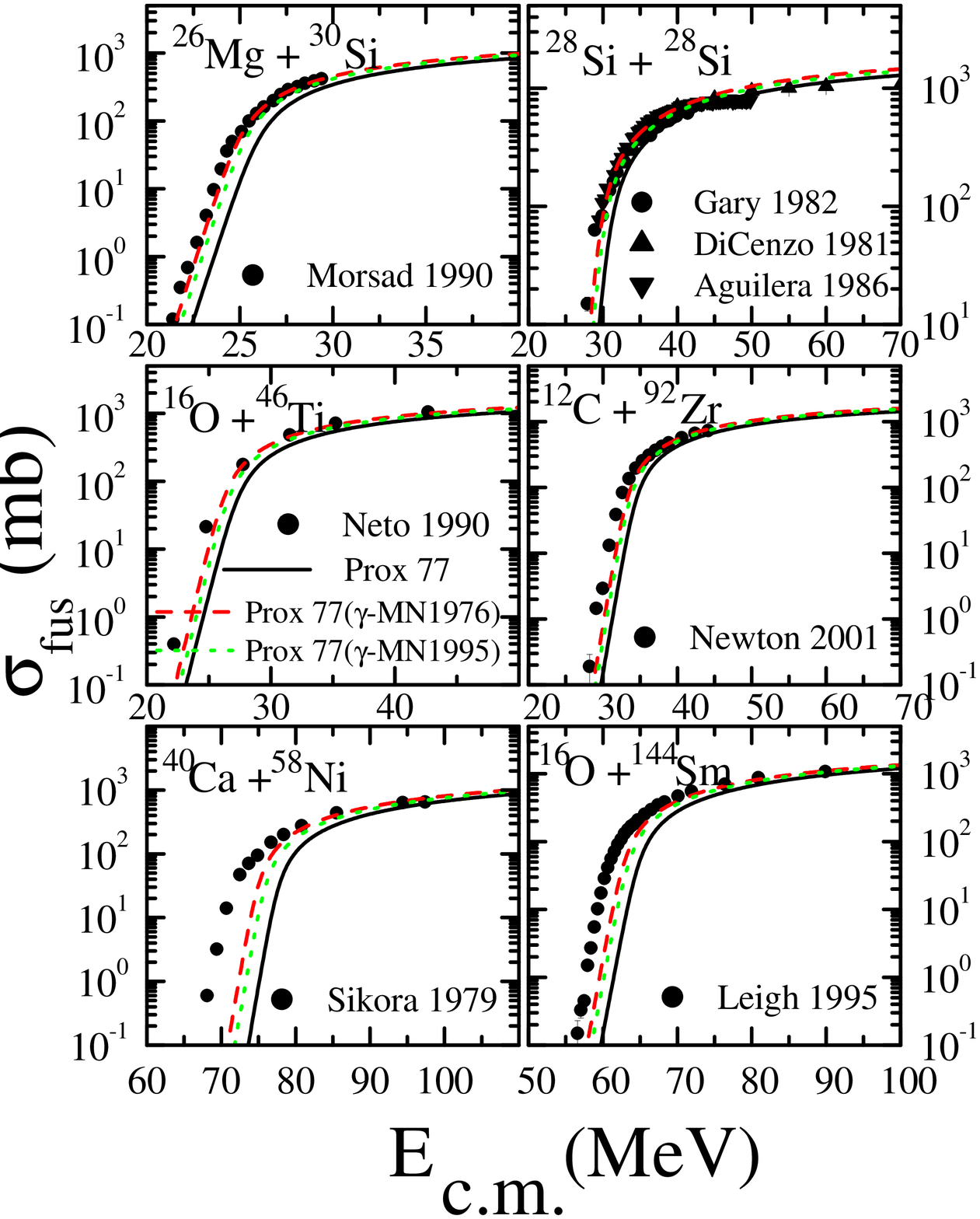}% Here is how to import EPS art
\vskip -0.7 cm \caption {(Color online) The fusion cross-sections
$\sigma_{fus}$ (mb) as a function of center-of-mass energy
$E_{c.m.}$ using different versions of $\gamma$ in Prox 77. The
original Prox 77 is also shown for comparison. The experimental
data are from Morsad 1990~\cite{Morsad90}, Gary
1982~\cite{Gary82}, DiCenzo 1981~\cite{sb81}, Aguilera
1986~\cite{Aguilera86}, Neto 1990~\cite{Neto90}, Newton
2001~\cite{newton01}, Sikora 1979~\cite{sikora79} and Leigh
1995~\cite{leigh95}.}\label{fig4}
% \vskip -0.6 cm
\end{figure}
%%%%%%%%%%%%%%%%%%%%%%%%%%%%%%%%%%%%%%%%%%%%%%%%%%%%%%%%%%%%%%%%%%%%%%%%%%%


\begin{thebibliography}{0}
\bibitem{blocki77} J. Blocki, J. Randrup, W. J. \'Swi\c{a}tecki, and C. F. Tsang, Ann. Phys. (N.Y.) \textbf{105,} 427 (1977).
\bibitem{wr94} W. Reisdorf, J. Phys. G: Nucl. Part. Phys. \textbf{20,} 1297 (1994).
\bibitem{ms2000} W. D. Myers and W. J. \'Swi\c{a}tecki, Phys. Rev. C \textbf{62,} 044610 (2000).
\bibitem{rkp05} R. K. Puri and N. K. Dhiman,  Eur. Phys. J. A {\bf23,}
429 (2005); R. Arora, R. K. Puri, and R. K. Gupta, {\it ibid.}
{\bf 8,} 103 2000; R. K. Puri, P. Chattopadhyay, and R. K. Gupta,
Phys. Rev. C {\bf 43,} 315 (1991).
\bibitem{aw95} A. Winther, Nucl. Phys. \textbf{A594,} 203 (1995).
\bibitem{ms66} W. D. Myers and W. J. \'Swi\c{a}tecki, Nucl. Phys. \textbf{81,} 1 (1966).
\bibitem{ms67} W. D. Myers and W. J. \'Swi\c{a}tecki, Ark. Fys. \textbf{36,} 343 (1967).
\bibitem{mn76} P. M\"oller and J. R. Nix, Nucl. Phys. \textbf{A272,} 502 (1976).
\bibitem{mn95} P. M\"oller, J. R. Nix, W. D. Myers, and W. J. \'Swi\c{a}tecki, At. Data Nucl. Data  Tables
\textbf{59,} 185 (1995).
\bibitem{mn81} P. M\"oller and J. R. Nix, Nucl. Phys. \textbf{A361,} 117 (1981).
\bibitem{rr84} G Royer and B Remaud, J. Phys. G: Nucl. Part. Phys. \textbf{10,} 1057 (1984).
\bibitem{pd03} K. Pomorski and J. Dudek, Phys. Rev. C \textbf{67,} 044316 (2003).
\bibitem{bn94} B. Nerlo-Pomorska and K. Pomorski, Z. Phys. A \textbf{348,} 169
(1994).
\bibitem{gr09} G. Royer and R. Rousseau,   Eur. Phys. J. A \textbf{42,} 541
(2009).
\bibitem{Vaz81}  V. Tripathi  \emph{et al.}, Phys. Rev. C \textbf{65},
014614 (2001); S. Sinha, M. R. Pahlavani, R. Varma, R. K.
Choudhury, B. K. Nayak, and A. Saxena, {\it ibid.} \textbf{64},
024607 (2001); I. Padron \emph{et al.}, {\it ibid.} \textbf{66},
044608 (2002); Z. H. Liu \emph{et al.}, Eur. Phys. J. \textbf{A
26}, 73 (2005); J. Skalski,  Phys. Rev. C {\bf 76,} 044603 (2007);
S. Mitsuoka, H. Ikezoe, K. Nishio, K. Tsuruta, S. C. Jeong, and Y.
Watanabe, Phys. Rev. Lett. \textbf{99}, 182701 (2007); A. M.
Stefanini \emph{et al.}, Phys. Rev. C \textbf{78}, 044607 (2008);
K. Washiyama and D. Lacroix, {\it ibid.} {\bf 78,} 024610 (2008);
C. J. Lin, H. M. Jia, H. Q. Zhang, F. Yang, X. X. Xu, F. Jia, Z.
H. Liu, and K. Hagino, {\it ibid.}. \textbf{ 79}, 064603 (2009);
A. M. Stefanini, \emph{et al.}, Phys. Lett. \textbf{B 679}, 95
(2009).
\bibitem{Morsad90} A. Morsad,J. J. Kolata, R. J. Tighe, X. J. Kong, E. F. Aguilera, and J. J. Vega, Phys. Rev. C \textbf{41}, 988 {1990}
 \bibitem{Gary82} S. Gary and C. Volant, Phys. Rev. C {\bf 25,} 1877
(1982).
 \bibitem{sb81} S. B. Dicenzo and J. F. Petersen, and R. R. Betts, Phys. Rev. C {\bf 23,}
 2561 (1981).
\bibitem{Aguilera86} E. F. Aguilera, J. J. Kolata, P. A. DeYoung, and J. J. Vega, Phys. Rev. C \textbf{33},  1961 (1986).
\bibitem{Neto90} R. Liguori Neto, J. C. Acquadro, P. R. S. Gomes, A. Szanto de Toledo, C. F. Tenreiro, E. Crema, N. C. Filho, and M. M. Coimbra, Nucl. Phys. \textbf{A512}, 333  (1990).
\bibitem{newton01} J. O. Newton, C. R. Morton, M. Dasgupta, J. R. Leigh, J. C. Mein, D. J. Hinde, H. Timmers, and K. Hagino, Phys. Rev. C \textbf{64}, 064608 (2001).
 \bibitem{sikora79}  B. Sikora, J. Bisplinghoff, W. Scobel, M. Beckerman, and M. Blann, Phys. Rev. C {\bf 20,}
 2219 (1979).
 \bibitem{leigh95} J. R. Leigh \emph{et al.}, Phys. Rev. C \textbf{52}, 3151 (1995).
 \bibitem{wg72} C. Y. Wong,  Phys. Lett. {\bf B42,} 186 (1972);
 C. Y. Wong,   Phys. Rev. Lett. {\bf 31,} 766 (1973).
%%%%%%%%%%%%%%%%%%%%%%%%%%%%%%%%%%%%%%%%%%%%%%%%%%Barrier fomula parametrization as we have given %%%%%%%%%%%%%%%%%%%%%%%%%%%%%%%%%%%%%%%%%%%%%%%%%%%%%%%%%

\end{thebibliography}
\end{document}